\documentclass[prb,preprintnumbers,amsmath,amssymb,floatfix]{revtex4}

\usepackage{graphicx}
\usepackage{subfigure}
\usepackage{dcolumn}

\begin{document}

\title{Quantum theory of electronic double-slit diffraction}
\author{Xiang-Yao Wu$^{a}$ \footnote{E-mail: wuxy2066@163.com}, Bai-Jun Zhang$^{a}$,
Xiao-Jing Liu$^{a}$,\\ Li Wang$^{a}$, Bing Liu$^{a}$, Xi-Hui
Fan$^{b}$ and Yi-Qing Guo$^{c}$}
\affiliation{a.Institute of Physics, Jilin Normal University, Siping 136000, China\\
b. Department of Physics, Qufu Normal University, Qufu 273165, China\\
c. Institute of High Energy Physics, P. O. Box 918(3), Beijing
100049, China}

\begin{abstract}
The phenomena of electron, neutron, atomic and molecular
diffraction have been studied by many experiments, and these
experiments are explained by some theoretical works. In this
paper, we study electronic double-slit diffraction with quantum
mechanical approach. We can obtain the results: (1) When the slit
width $a$ is in the range of $3\lambda\sim 50\lambda$ we can
obtain the obvious diffraction patterns. (2) when the ratio of
$\frac{d+a}{a}=n (n=1, 2, 3,\cdot\cdot\cdot)$, order $2n, 3n,
4n,\cdot\cdot\cdot$ are missing in diffraction pattern. (3)When
the ratio of $\frac{d+a}{a}\neq n (n=1, 2, 3,\cdot\cdot\cdot)$,
there isn't missing order in diffraction pattern. (4) We also find
a new quantum mechanics effect that the slit thickness $c$ has a
large affect to the electronic diffraction patterns. We think all
the predictions in our work can be tested by the electronic
double-slit diffraction experiment.

\end{abstract}
\maketitle

{\bf 1. Introduction} \vskip 5pt

The wave nature of subatomic particle elections and neutrons was
postulated by de Broglie in 1923 and this idea can explain many
diffraction experiments. The matter-wave diffraction has become a
large field of interest throughout the last year, and it is extend
to atom, more massive, complex objects, like large molecules
$I_{2}$, $C_{60}$ and $C_{70}$, which were found in
experiments\cite{s1}\cite{s2}\cite{s3}\cite{s4}\cite{s5}. In such
experiments, the incoming atoms or molecules usually can be
described by plane wave. As well known, the classical optics with
its standard wave-theoretical methods and approximations, in
particular those of Huygens and Kirchhoff, has been successfully
applied to classical optics, and has yielded good agreement with
many experiments. This simple wave-optical approach gives a
description of matter wave diffraction also\cite{s6}\cite{s7}.
However, matter-wave interference and diffraction are quantum
phenomena, and its fully description needs quantum mechanical
approach. In this work, we study the double-slit diffraction of
electron with quantum mechanical approach. In the view of quantum
mechanics, the electron has the nature of wave, and the wave is
described by wave function $\psi(\vec{r},t)$, which can be
calculated with Schr\"{o}dinger's wave equation. The wave function
$\psi(\vec{r},t)$ has statistical meaning, i.e.,
$\mid\psi(\vec{r},t)\mid^{2}$ can be explained as the particle's
probability density at the definite position. For double-slit
diffraction, if we can calculate the electronic wave function
$\psi(\vec{r},t)$ distributing on display screen, then we can
obtain the diffraction intensity for the double-slit, since the
diffraction intensity is directly  proportional to
$\mid\psi(\vec{r},t)\mid^{2}$. In the double-slit diffraction, the
electron wave functions can be divided into three areas. The first
area is the incoming area, the electronic wave function is a plane
wave. The second area is the double-slit area, where the
electronic wave function can be calculated by Schr\"{o}dinger's
wave equation. The third area is the diffraction area, where the
electronic wave function can be obtained by the Kirchhoff's law.
In the following, we will calculate these wave functions.

The paper is organized as follows. In section 2 we calculate the
electronic wave function in the double-slit with quantum
mechanical approach. In section 3 we calculate the electronic wave
function in diffraction area with the Kirchhoff's law. Section 4
is numerical analysis, Section 5 is a summary of results and
conclusion. \vskip 5pt
\newpage
 \setlength{\unitlength}{0.1in}
 \begin{center}
\begin{figure}
\begin{picture}(100,15)
 \put(30,4){\vector(1,0){13}}
 \put(30,4){\vector(0,1){10}}
 \put(30,4){\vector(2,1){11}}
 \put(26,2){\line(1,0){18}}
 \put(26,2){\line(0,1){13}}
 \put(44,2){\line(0,1){13}}
 \put(26,15){\line(1,0){18}}
 \put(30,8){\line(1,0){2}}
 \put(32,4){\line(0,1){4}}
\put(36,4){\line(0,1){4}}
 \put(36,8){\line(1,0){2}}
 \put(38,4){\line(0,1){4}}
 \put(42,2.5){\makebox(2,1)[l]{$y$}}
 \put(30,13){\makebox(2,1)[c]{$x$}}
 \put(30,2.6){\makebox(2,1)[l]{$o$}}
 \put(40,8){\makebox(2,1)[c]{$z$}}
 \put(28,6){\makebox(2,1)[c]{$b$}}
 \put(30,8){\makebox(2,1)[c]{$a$}}

\put(32,7){\line(1,0){1}}
\put(35,7){\line(1,0){1}}
 \put(33.5,6.5){\makebox(2,1)[l]{$d$}}
\end{picture}
\caption{The double-slit geometry, the width $a$, the length $b$
and the two slit distance $d$. } \label{moment}
\end{figure}
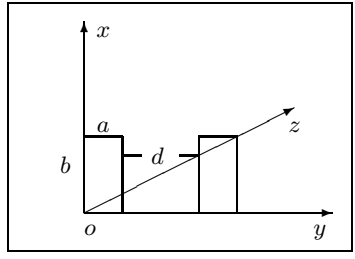
\end{center}
{\bf 2. Quantum theory of electron diffraction in double-slit }
\vskip 10pt

 In an infinite plane, we consider a double-slit, its
width $a$, length $b$ and the two slit distance $d$ are shown in
FIG. 1. The $x$ axis is along the slit length $b$ and the $y$ axis
is along the slit width $a$. In the following, we calculate the
electron wave function in the first single-slit (left slit) with
Schrodinger equation, and the electron wave function of the second
single-slit(right slit) can be obtained easily. At a time $t$, we
suppose that the incoming plane wave travels along the $z$ axis.
It is
\begin{equation}
\Psi_{1}(z, t)=Ae^{\frac{i}{\hbar}(pz-Et)},
\end{equation}
where $A$ is a constant.

The potential in the single-slit is
\begin{eqnarray}
V(x,y,z)&=&0  \hspace{0.3in}0\leq x\leq b, 0\leq y\leq a, 0\leq z\leq c\nonumber\\
       &=&\infty \hspace{0.3in} otherwise,
\end{eqnarray}
where $c$ is the thickness of the single-slit. The time-dependent
and time-independent Schrodinger equations are
\begin{equation}
i\hbar\frac{\partial}{\partial
t}\psi(\vec{r},t)=-\frac{\hbar^{2}}{2M}(\frac{\partial^{2}}{\partial
x^{2}}+\frac{\partial^{2}}{\partial
y^{2}}+\frac{\partial^{2}}{\partial z^{2}})\psi(\vec{r},t),
\end{equation}
\begin{equation}
\frac{\partial^{2}\psi(\vec{r})}{\partial
x^{2}}+\frac{\partial^{2}\psi(\vec{r})}{\partial
y^{2}}+\frac{\partial^{2}\psi(\vec{r})}{\partial
z^{2}}+\frac{2ME}{\hbar^{2}}\psi(\vec{r})=0,
\end{equation}
where $M(E)$ is the mass(energy) of the electron. In Eq. (4), the
wave function $\psi(x,y,z)$ satisfies boundary conditions
\begin{equation}
\psi(0,y,z)=\psi(b,y,z)=0,
\end{equation}
\begin{equation}
\psi(x,0,z)=\psi(x,a,z)=0,
\end{equation}
The partial differential equation (4) can be solved by the method
of separation of variable. By writing

\begin{equation}
\psi(x,y,z)=X(x)Y(y)Z(z),
\end{equation}
Eq. (4) becomes
\begin{equation}
\frac{1}{X}\frac{d^{2}X}{d x^{2}}+\frac{1}{Y}\frac{d^{2}Y}{d
y^{2}}+\frac{1}{Z}\frac{d^{2}Z}{d z^{2}}+\frac{2ME}{\hbar^{2}}=0,
\end{equation}
and Eq. (8) can be written into the following three equations
\begin{equation}
\frac{1}{X}\frac{d^{2}X}{d x^{2}}+\lambda_{1}=0,
\end{equation}
\begin{equation}
\frac{1}{Y}\frac{d^{2}Y}{d y^{2}}+\lambda_{2}=0,
\end{equation}
\begin{equation}
\frac{1}{Z}\frac{d^{2}Z}{d x^{2}}+\lambda_{3}=0,
\end{equation}
where $\lambda_{1}$, $\lambda_{2}$ and $\lambda_{3}$ are
constants,which satisfy
\begin{equation}
\frac{2ME}{\hbar^2}=\lambda_{1}+\lambda_{2}+\lambda_{3}.
\end{equation}
From Eq. (5) and (6), with $X(x)$ and $Y(y)$ satisfying the
boundary conditions
\begin{eqnarray}
X(0)=X(b)=0  \nonumber\\
Y(0)=Y(a)=0,
\end{eqnarray}
we can obtain the equations of $X(x)$ and $Y(y)$
\begin{eqnarray}
&&\frac{d^{2}X}{d x^{2}}+\lambda_{1}X=0 \nonumber\\
&&X(0)=0 \nonumber\\
&&X(b)=0,
\end{eqnarray}
and
\begin{eqnarray}
&&\frac{d^{2}Y}{d y^{2}}+\lambda_{2}Y=0\nonumber\\
&&Y(0)=0\nonumber\\
&&Y(a)=0,
\end{eqnarray}
their eigenfunctions and eigenvalues are
\begin{eqnarray}
&&X_{n}(x)=A_{n}\sin{\frac{n\pi}{b}x}
\hspace{0.3in}(n=1,2,\cdots)\nonumber\\&&
\lambda_{1}=(\frac{n\pi}{b})^{2},
\end{eqnarray}
and
\begin{eqnarray}
&&Y_{m}(y)=B_{m}\sin{\frac{m\pi}{a}y}
\hspace{0.3in}(m=1,2,\cdots)\nonumber\\&&
\lambda_{2}=(\frac{m\pi}{a})^{2}.
\end{eqnarray}
The solution of Eq. (11) is
\begin{equation}
Z_{mn}(z)=C_{mn}e^{i
\sqrt{\frac{2ME}{\hbar^{2}}-\frac{n^{2}\pi^{2}}{b^{2}}-\frac{m^{2}\pi^{2}}{a^{2}}}
 z},
\end{equation}
and the particular solution of the wave equation (4) is
\begin{eqnarray}
\psi_{mn}&=&X_{n}(x)Y_{m}(y)Z_{mn}(z) \nonumber\\
&=&A_{n}B_{m}C_{mn}\sin{\frac{n\pi x}{b}}\sin{\frac{m\pi
y}{a}}e^{i\sqrt{\frac{2ME}{\hbar^{2}}-\frac{n^{2}\pi^{2}}{b^{2}}-\frac{m^{2}\pi^{2}}{a^{2}}}z} \nonumber\\
&=&D_{mn}\sin{\frac{n\pi x}{b}}\sin{\frac{m\pi
y}{a}}e^{i\sqrt{\frac{2ME}{\hbar^{2}}-\frac{n^{2}\pi^{2}}{b^{2}}-\frac{m^{2\pi^{2}}}{a^{2}}}z}.
\end{eqnarray}
The time-dependent particular solution of Eq. (3) is
\begin{equation}
\psi_{mn}(x,y,z,t)=\psi_{mn}(x,y,z)e^{-\frac{i}{\hbar}Et}.
\end{equation}
The general solution of Eq. (3) is
\begin{eqnarray}
\psi_{2}(x,y,z,t)&=&\sum_{mn}\psi_{mn}(x,y,z,t) \nonumber\\
&=&\sum_{mn}D_{mn}\sin{\frac{n\pi x}{b}}\sin{\frac{m\pi
y}{a}}e^{i\sqrt{\frac{2ME}{\hbar^{2}}-\frac{n^{2}\pi^{2}}{b^{2}}-\frac{m^{2}\pi^{2}}{a^{2}}}z}e^{-\frac{i}{\hbar}Et}.
\end{eqnarray}
Eq. (21) is the electronic wave function in the first single-slit.
Since the wave functions are continuous at $z=0$, we have
\begin{equation}
\psi_{1}(x,y,z,t)\mid_{z=0}=\psi_{2}(x,y,z,t)\mid_{z=0},
\end{equation}
from Eq. (22), we obtain
\begin{equation}
\sum_{mn}D_{mn}\sin{\frac{n\pi x}{b}}\sin{\frac{m\pi y}{a}}=A,
\end{equation}
where $D_{mn}$ is a coefficient, which is
\begin{eqnarray}
D_{mn}&=&\frac{4}{a b}\int^{a}_{0}\int^{b}_{0}A\sin{\frac{n\pi
\xi}{b}}\sin{\frac{m\pi \eta}{a}}d\xi d\eta \nonumber\\
&=&\frac{16A}{mn\pi^{2}} \hspace{0.6in} m,n, odd  \nonumber\\
&=&0 \hspace{0.9in} otherwise,
\end{eqnarray}
substituting Eq. (24) into (21), we can obtain the electronic wave
function in the first single-slit.

\begin{eqnarray}
\psi_{2}(x,y,z,t)&=&\sum_{m,n=0}^{\infty}\frac{16A}{(2m+1)(2n+1)\pi^{2}}\sin{\frac{(2n+1)\pi
x}{b}}\sin{\frac{(2m+1)\pi y}{a}} \nonumber\\&&
e^{i\sqrt{\frac{2ME}{\hbar^{2}}-\frac{(2n+1)^{2}\pi^{2}}{b^{2}}
-\frac{(2m+1)^{2}\pi^{2}}{a^{2}}}z}e^{-\frac{i}{\hbar}Et}.
\end{eqnarray}

The electron wave function in the second single-slit can be
obtained by making the coordinate translation:
\begin{eqnarray}
x^{'}&=&x\nonumber\\
y^{'}&=&y-(a+d)\nonumber\\
z^{'}&=&z,
\end{eqnarray}
on substituting Eq. (26) into (25), we can obtain the electron
wave function $\psi_{3}(x,y,z,t)$ in the second single-slit
\begin{eqnarray}
\psi_{3}(x,y,z,t)&=&\sum_{m,n=0}^{\infty}\frac{16A}{(2m+1)(2n+1)\pi^{2}}\sin{\frac{(2n+1)\pi
x}{b}}\sin{\frac{(2m+1)\pi (y-(a+d))}{a}} \nonumber\\&&
e^{i\sqrt{\frac{2ME}{\hbar^{2}}-\frac{(2n+1)^{2}\pi^{2}}{b^{2}}
-\frac{(2m+1)^{2}\pi^{2}}{a^{2}}}z}e^{-\frac{i}{\hbar}Et}.
\end{eqnarray}
\vskip 8pt

{\bf 3. The wave function of electron diffraction } \vskip 8pt
With the Kirchhoff's law, we can calculate the electron wave
function in the diffraction area. It can be calculated by the
formula\cite{s8}
\begin{equation}
\psi_{out}(\vec{r},t)=-\frac{1}{4\pi}\int_{s}\frac{e^{ikr}}{r}\overrightarrow{n}\cdot[\bigtriangledown^{'}\psi_{in}
+(ik-\frac{1}{r})\frac{\overrightarrow{r}}{r}\psi_{in}]ds,
\end{equation}
where $\psi_{out}(\vec{r},t)$ is diffraction wave function on
display screen, $\psi_{in}(\vec{r},t)$ is the wave function of
slit surface (z=c) and $s$ is the area of the aperture or slit.
The Kirchhoff formula (28) is approximate, since it neglects the
effect of diffraction aperture or slit on the incoming wave
$\psi_{in}(\vec{r},t)$. However, when the diffraction aperture or
slit is larger than the electron wave
length the effect can be neglected.\\
For the double-slit diffraction, Eq. (28) becomes
\begin{eqnarray}
\psi_{out}(\vec{r},t)=&&-\frac{1}{4\pi}\int_{s_{1}}\frac{e^{ikr}}{r}\overrightarrow{n}\cdot[\bigtriangledown^{'}\psi_{2}
+(ik-\frac{1}{r})\frac{\overrightarrow{r}}{r}\psi_{2}]ds
\nonumber\\&&
-\frac{1}{4\pi}\int_{s_{2}}\frac{e^{ikr}}{r}\overrightarrow{n}\cdot[\bigtriangledown^{'}\psi_{3}
+(ik-\frac{1}{r})\frac{\overrightarrow{r}}{r}\psi_{3}]ds.
\end{eqnarray}
In Eq. (29), the first and second terms are corresponding to the
diffraction wave functions of the first slit and the second slit.

   In the following, we firstly calculate the diffraction wave
function of the first slit, it is
\begin{equation}
\psi_{out_{1}}(\vec{r},t)=-\frac{1}{4\pi}\int_{s_{1}}\frac{e^{ikr}}{r}\overrightarrow{n}\cdot[\bigtriangledown^{'}\psi_{2}
+(ik-\frac{1}{r})\frac{\overrightarrow{r}}{r}\psi_{2}]ds,
\end{equation}
The diffraction area is shown in FIG. 2, where
$k=\sqrt{\frac{2ME}{\hbar^{2}}}$, $s_{1}$ is the area of the first
single-slit, $\overrightarrow{r}^{'}$ is the position of a point
on the surface (z=c), $P$ is an arbitrary point in the diffraction
area, and $\overrightarrow{n}$ is a unit vector, which is normal
to the surface of the slit.
\setlength{\unitlength}{0.1in}
 \begin{center}
\begin{figure}
\begin{picture}(100,10)
 \put(26,5){\vector(1,0){3}}
 \put(26,5){\vector(0,1){2.2}}
 \put(26,5){\vector(2,1){5}}

 \put(24,1){\line(1,0){2}}
 \put(24,1){\line(0,1){4}}
 \put(26,1){\line(0,1){4}}
 \put(24,5){\line(1,0){2}}

 \put(24,10){\line(1,0){2}}
 \put(24,10){\line(0,1){4}}
 \put(26,10){\line(0,1){4}}
 \put(24,14){\line(1,0){2}}

 \put(26,7){\line(3,1){11.5}}
 \put(26,7){\vector(3,1){5}}
 \put(26,5){\line(2,1){11.5}}
 \put(27.5,3.2){\makebox(2,1)[l]{$\overrightarrow{n}$}}
 \put(24,7){\makebox(2,1)[c]{$\overrightarrow{r}^{\prime}$}}
 \put(25,5){\makebox(2,1)[l]{$o$}}
 \put(31,5.5){\makebox(2,1)[c]{$\overrightarrow{R}$}}
 \put(30,9){\makebox(2,1)[c]{$\overrightarrow{r}$}}
 \put(38.5,10.5){\makebox(2,1)[l]{$P$}}
 \put(25,-0.5){\makebox(2,1)[l]{$c$}}
\end{picture}
\caption{The diffraction area of single-slit} \label{moment}
\end{figure}
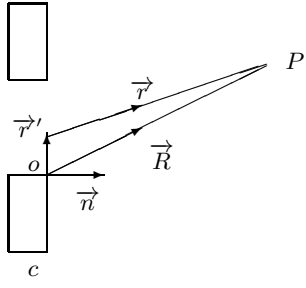
\end{center}

From FIG. 2, we have
\begin{eqnarray}
r&=&R-\frac{\overrightarrow{R}}{R}\cdot\overrightarrow{r}^{'}\nonumber\\
&\approx&R-\frac{\overrightarrow{r}}{r}\cdot\overrightarrow{r}^{'}\nonumber\\
 &=&R-\frac{\overrightarrow{k_{2}}}{k}\cdot\overrightarrow{r}^{'},
\end{eqnarray}
then,
\begin{eqnarray}
\frac{e^{ikr}}{r}&=&\frac{e^{ik(R-\frac{\overrightarrow{r}}{r}\cdot\overrightarrow{r}^{'})}}
{R-\frac{\overrightarrow{r}}{r}\cdot\overrightarrow{r}^{'}}\nonumber\\
&=&\frac{e^{ikR}e^{-i\overrightarrow{k_{2}}\cdot\overrightarrow{r}^{'}}}
{R-\frac{\overrightarrow{r}}{r}\cdot\overrightarrow{r}^{'}}\nonumber\\
&\approx&\frac{{e^{ikR}e^{-i\overrightarrow{k_{2}}\cdot\overrightarrow{r}^{'}}}}{R}
\hspace{0.3in}(|\overrightarrow{r}^{'}|\ll R),
\end{eqnarray}
with $\vec{K_{2}}=K\frac{\vec{r}}{r}$. Substituting Eq. (31) and
(32) into (30), one can obtain
\begin{equation}
\psi_{out_{1}}(\vec{r},t)=-\frac{e^{ikR}}{4\pi
R}\int_{s_{0}}e^{-i\overrightarrow{k_{2}}\cdot
\overrightarrow{r}^{'}}\overrightarrow{n}\cdot
[\bigtriangledown^{'}\psi_{2}(x^{'},y^{'},z^{'})+
(i\overrightarrow{k}_{2}-\frac{\overrightarrow{R}}{R^{2}})\psi_{2}(x^{'},y^{'},z^{'})]ds^{'}.
\end{equation}
In Eq. (33), the term
\begin{eqnarray}
\overrightarrow{n}\cdot
\nabla^{'}\psi_{2}(x^{'},y^{'},z^{'})|_{z=c}&=&n_{x}\frac{\partial
\psi_{2}(\overrightarrow{r}^{'})}{\partial
x^{'}}+n_{y}\frac{\partial
\psi_{2}(\overrightarrow{r}^{'})}{\partial
y^{'}}+n_{z}\frac{\partial
\psi_{2}(\overrightarrow{r}^{'})}{\partial z^{'}}\nonumber\\
&=&n_{z}\frac{\partial
\psi_{2}(\overrightarrow{r}^{'})}{\partial z^{'}}\nonumber\\
&=&\sum_{m=0}^{\infty}\sum_{n=0}^{\infty}\frac{16A}{(2m+1)(2n+1)\pi^{2}}\nonumber\\&&
i\sqrt{\frac{2ME}{\hbar^{2}}-(\frac{(2n+1)\pi}{b})^{2}-(\frac{(2m+1)\pi}{a})^{2}}\nonumber\\&&
e^{i\sqrt{\frac{2ME}{\hbar^{2}}-(\frac{(2n+1)\pi}{b})^{2}-(\frac{(2m+1)\pi}{a})^{2}}\cdot
c}\nonumber\\&& \sin \frac{(2n+1)\pi}{b}x^{'}\sin
\frac{(2m+1)\pi}{a}y^{'},
\end{eqnarray}
then Eq. (33) is
\begin{eqnarray}
\psi_{out_{1}}(\vec{r},t)&=&-\frac{e^{ikR}}{4\pi
R}e^{-\frac{i}{\hbar}Et}\int_{s_{0}}e^{-i\overrightarrow{k_{2}}\cdot
\overrightarrow{r}^{'}}\sum_{m=0}^{\infty}\sum_{n=0}^{\infty}\frac{16A}{(2m+1)(2n+1)\pi^{2}}\nonumber\\&&
e^{i\sqrt{\frac{2ME}{\hbar^{2}}-(\frac{(2n+1)\pi}{b})^{2}-(\frac{(2m+1)\pi}{a})^{2}}\cdot
c} \sin \frac{(2n+1)\pi}{b}x^{'}\sin
\frac{(2m+1)\pi}{a}y^{'}\nonumber\\&&
[i\sqrt{\frac{2ME}{\hbar^{2}}-(\frac{(2n+1)\pi}{b})^{2}-(\frac{(2m+1)\pi}{a})^{2}}+i\overrightarrow{n}\cdot
\overrightarrow{k_{2}}-\frac{\overrightarrow{n}\cdot
\overrightarrow{R}}{R^{2}}]dx^{'}dy^{'}.
\end{eqnarray}
Assume that the angle between $\overrightarrow{k_{2}}$ and $x$
axis ($y$ axis) is $\frac{\pi}{2}-\alpha$ ($\frac{\pi}{2}-\beta$),
and $\alpha (\beta)$ is the angle between $\overrightarrow{k_{2}}$
and the surface of $yz$ ($xz$), then we have
\begin{eqnarray}
k_{2x}=k\sin \alpha,\hspace{0.3in} k_{2y}=k\sin \beta,
\end{eqnarray}
\begin{eqnarray}
\overrightarrow{n}\cdot \overrightarrow{k_{2}}=k\cos \theta,
\end{eqnarray}
where $\theta$ is the angle between $\overrightarrow{k_{2}}$ and
$z$ axis, and the angles $\theta$, $\alpha$, $\beta$ satisfy the
equation
\begin{equation}
\cos^{2}\theta+\cos^{2}(\frac{\pi}{2}-\alpha)+\cos^{2}(\frac{\pi}{2}-\beta)=1.
\end{equation}
Substituting Eq. (36) - (38) into (35) gives

\begin{eqnarray}
\psi_{out_{1}}(x,y,z,t)&=&-\frac{e^{ikR}}{4\pi
R}e^{-\frac{i}{\hbar}Et}\sum_{m=0}^{\infty}\sum_{n=0}^{\infty}\frac{16A}{(2m+1)(2n+1)\pi^2}
e^{i\sqrt{\frac{2ME}{\hbar^{2}}-(\frac{(2n+1)\pi}{b})^{2}-(\frac{(2m+1)\pi}{a})^{2}}\cdot
c}\nonumber\\&&
[i\sqrt{\frac{2ME}{\hbar^{2}}-(\frac{(2n+1)\pi}{b})^{2}-(\frac{(2m+1)\pi}{a})^{2}}+(ik-\frac{1}{R})
\sqrt{\cos^{2}\alpha-\sin^{2}\beta}]\nonumber\\&&
\int^{b}_{0}e^{-ik\sin\alpha\cdot
x^{'}}\sin\frac{(2n+1)\pi}{b}x^{'}dx^{'}\int^{a}_{0}e^{-ik\sin\beta\cdot
y^{'}} \sin \frac{(2m+1)\pi}{a}y^{'}dy^{'}.
\end{eqnarray}
Eq. (39) is the diffraction wave function of the first slit.
Obviously, the diffraction wave function of the second slit is
\begin{eqnarray}
\psi_{out_{2}}&&(x,y,z,t)=-\frac{e^{ikR}}{4\pi
R}e^{-\frac{i}{\hbar}Et}\sum_{m=0}^{\infty}\sum_{n=0}^{\infty}\frac{16A}{(2m+1)(2n+1)\pi^2}
e^{i\sqrt{\frac{2ME}{\hbar^{2}}-(\frac{(2n+1)\pi}{b})^{2}-(\frac{(2m+1)\pi}{a})^{2}}\cdot
c}\nonumber\\&&
[i\sqrt{\frac{2ME}{\hbar^{2}}-(\frac{(2n+1)\pi}{b})^{2}-(\frac{(2m+1)\pi}{a})^{2}}+(ik-\frac{1}{R})
\sqrt{\cos^{2}\alpha-\sin^{2}\beta}]\nonumber\\&&
\int^{b}_{0}e^{-ik\sin\alpha\cdot
x^{'}}\sin\frac{(2n+1)\pi}{b}x^{'}dx^{'}\int^{2a+d}_{a+d}e^{-ik\sin\beta\cdot
y^{'}} \sin \frac{(2m+1)\pi}{a}(y^{'}-(a+d))dy^{'},
\end{eqnarray}
where $d$ is the two slit distance. The total diffraction wave
function for the double-slit is
\begin{eqnarray}
\psi_{out}(x,y,z,t)=\psi_{out_{1}}(x,y,z,t)+\psi_{out_{2}}(x,y,z,t)
\end{eqnarray}

From the diffraction wave function $\psi_{out}(x,y,z,t)$, we can
obtain the relative diffraction intensity $I$ on the display
screen, it is
\begin{equation}
I\propto|\psi_{out}(x,y,z,t)|^{2}.
\end{equation}
\vskip 8pt

{\bf 4. Numerical result} \vskip 8pt

In this section we present our numerical calculation of relative
diffraction intensity. The main input parameters are:
$M=9.11\times10^{-31}kg$, $R=1m$,  $A=10^{8}$, $\alpha=0.01rad$,
$E=0.001eV$ and the Planck constant $\hbar=1.055\times10^{-34}Js$.
We can obtain the relation between the diffraction angle $\beta$
and relative diffraction intensity $I$. In double-slit
diffraction, we can obtain the results: (1) When the ratio of
$\frac{d+a}{a}=n (n=1, 2, 3,\cdot\cdot\cdot)$, order $2n, 3n,
4n,\cdot\cdot\cdot$ are missing in diffraction pattern. (2)When
the ratio of $\frac{d+a}{a}\neq n (n=1, 2, 3,\cdot\cdot\cdot)$,
there isn't missing order in diffraction pattern. In FIG. 3 and
FIG. 4, we take $a=\lambda$, $b=1000\lambda$ and $c=\lambda$, the
diffraction patterns are not obvious. In FIG. 3, the ratio of
$\frac{d+a}{a}=6$, the order 6 is missing. In FIG. 4, the ratio of
$\frac{d+a}{a}=6.5$, there isn't missing order. In FIG. 5, FIG. 6
and FIG. 7, we take $a=\lambda, b=1000\lambda, c=\lambda$, the
diffraction patterns are obvious, where $\lambda=\frac{2\pi
\hbar}{\sqrt{2ME}}$ is electronic wavelength. In FIG.5, the ratio
of $\frac{d+a}{a}=3$, the orders $3, 6, \cdot\cdot\cdot$ are
missing. In FIG. 6, the ratio of $\frac{d+a}{a}=3.4$, there isn't
missing order. In FIG. 7, the ratio of $\frac{d+a}{a}=6$, the
orders $6, 12, \cdot\cdot\cdot$ are missing. In FIG. 8, FIG. 9 and
FIG.10, the slit width $a$ are corresponding to $20\lambda$,
$30\lambda$ and $50\lambda$, their diffraction patterns are
obvious. In FIG. 8, FIG. 9 and FIG. 10, the ratio of
$\frac{d+a}{a}=3$, the orders $3, 6,\cdot\cdot\cdot$ are missing.
From FIG. 11 to FIG. 14, the slit thickness $c$ is corresponding
to $0$, $10\lambda$, $100\lambda$ and $1000\lambda$. We can find
that the thickness $c$ can make a large impact on the double-slit
diffraction pattern. When the slit thickness $c$ increases the
peak values of diffraction pattern increase also.

 \vskip 8pt
 \newpage
{\bf 5. Conclusion} \vskip 8pt

In conclusion, we studied the double-slit diffraction phenomenon
of electron with quantum mechanical approach. We give the relation
between diffraction angle and the relative diffraction intensity.
We find the following results: (1) When the slit width $a$ is in
the range of $3\lambda\sim 50\lambda$ we can obtain the obvious
diffraction patterns. (2) when the ratio of $\frac{d+a}{a}=n (n=1,
2, 3,\cdot\cdot\cdot)$, order $2n, 3n, 4n,\cdot\cdot\cdot$ are
missing in diffraction pattern. (3)When the ratio of
$\frac{d+a}{a}\neq n (n=1, 2, 3,\cdot\cdot\cdot)$, there isn't
missing order in diffraction pattern. (4) We also find a new
quantum mechanics effect that the slit thickness $c$ has a large
affect to the electronic diffraction patterns. We think all the
predictions in our work can be tested by the electronic
double-slit diffraction experiment.
 \vskip 10pt

{\bf  Acknowledgement} \vskip 8pt We are very grateful for the
valuable discussions with Yang Mao-Zhi.

\newpage

\begin{figure}[tbp]
\begin{picture}(25,0)
 \put(8,20){\makebox(2,1)[l]{$I$}}
 \put(19,3.5){\makebox(2,1)[c]{$\beta(rad)$}}
{\resizebox{5.5cm}{5.5cm}{\includegraphics{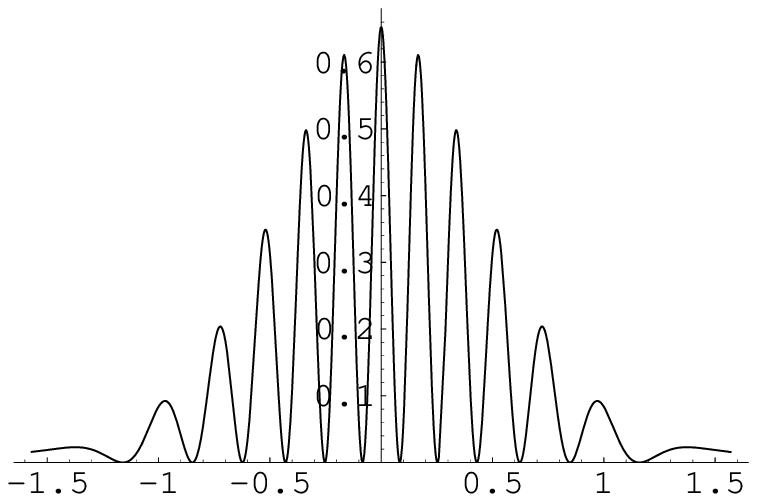}}}
 \put(12,20){\makebox(2,1)[l]{$I$}}
 \put(24,3.5){\makebox(2,1)[c]{$\beta(rad)$}}
\end{picture}
{\resizebox{5.5cm}{5.5cm}{\includegraphics{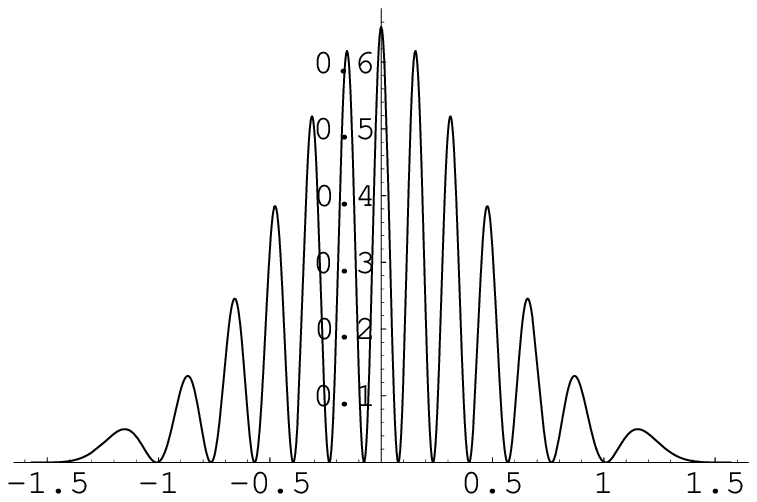}}} \vskip 5pt
{\hspace{0.1in}FIG. 3: The relation between $\beta$ and $I$
\hspace{0.35in} FIG.
4: The relation between $\beta$ and $I$ \\
 \hspace{0.3in}with $a=\lambda$,
$b=1000\lambda$, $c=\lambda$ and $d=5\lambda$. \hspace{0.4in} with
$a=\lambda$, $b=1000\lambda$, $c=\lambda$ and $d=5.5\lambda$.}
 \label{moment}
\end{figure}

\begin{figure}[tbp]
\begin{picture}(25,0)
 \put(8,20){\makebox(2,1)[l]{$I$}}
 \put(19,3.5){\makebox(2,1)[c]{$\beta(rad)$}}
{\resizebox{5.5cm}{5.5cm}{\includegraphics{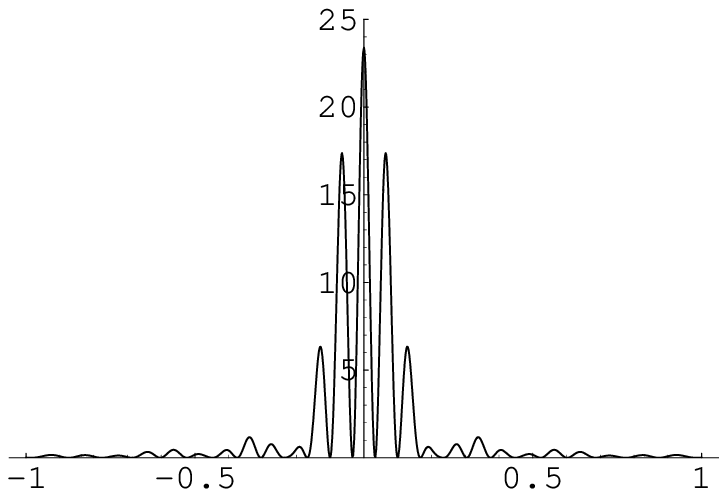}}}
 \put(12,20){\makebox(2,1)[l]{$I$}}
 \put(20,3.5){\makebox(2,1)[c]{$\beta(rad)$}}
\end{picture}
{\resizebox{5.5cm}{5.5cm}{\includegraphics{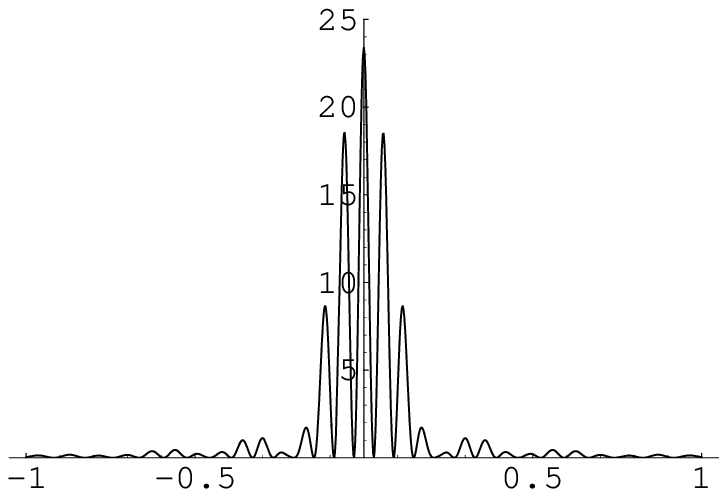}}} \vskip 5pt
{\hspace{0.1in}FIG. 5: The relation between $\beta$ and $I$
\hspace{0.35in} FIG.
6: The relation between $\beta$ and $I$ \\
 \hspace{0.3in}with $a=5\lambda$,
 $b=1000\lambda$, $c=\lambda$ and $d=10\lambda$. \hspace{0.4in}
with $a=5\lambda$, $b=1000\lambda$, $c=\lambda$ and
$d=12\lambda$.}
 \label{moment}
\end{figure}

\begin{figure}[tbp]
\begin{picture}(25,0)
 \put(8,20){\makebox(2,1)[l]{$I$}}
 \put(19,3.5){\makebox(2,1)[c]{$\beta(rad)$}}
{\resizebox{5.5cm}{5.5cm}{\includegraphics{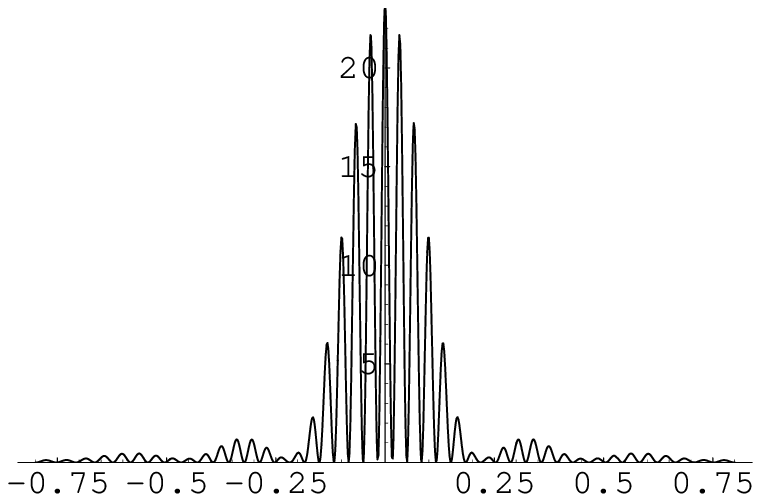}}}
 \put(12,20){\makebox(2,1)[l]{$I$}}
 \put(20,3.5){\makebox(2,1)[c]{$\beta(rad)$}}
\end{picture}
{\resizebox{5.5cm}{5.5cm}{\includegraphics{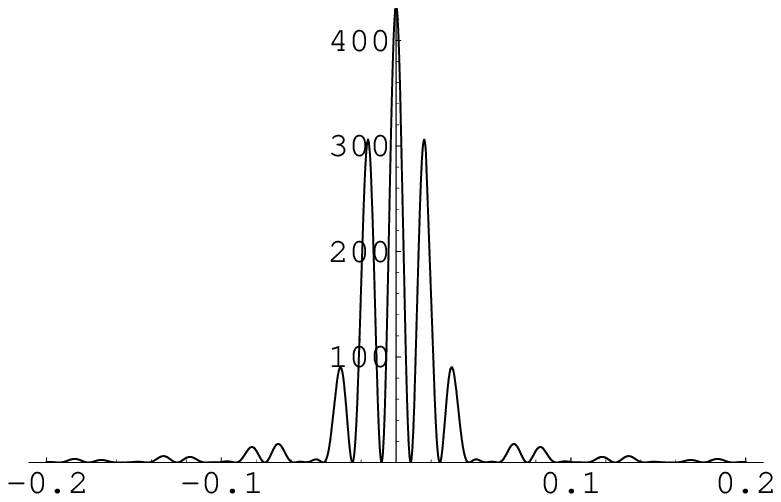}}} \vskip 5pt
{\hspace{0.1in}FIG. 7: The relation between $\beta$ and $I$ with
\hspace{0.35in} FIG.
8: The relation between $\beta$ and $I$ \\
 \hspace{0.3in} $a=5\lambda$,
$b=1000\lambda$, $c=\lambda$ and $d=25\lambda$. \hspace{0.4in}
with $a=20\lambda$ , $b=1000\lambda$, $c=\lambda$ and
$d=40\lambda$.}
 \label{moment}
\end{figure}

\begin{figure}[tbp]
\begin{picture}(25,0)
 \put(6,20){\makebox(2,1)[l]{$I$}}
 \put(19,3.5){\makebox(2,1)[c]{$\beta(rad)$}}
{\resizebox{5.5cm}{5.5cm}{\includegraphics{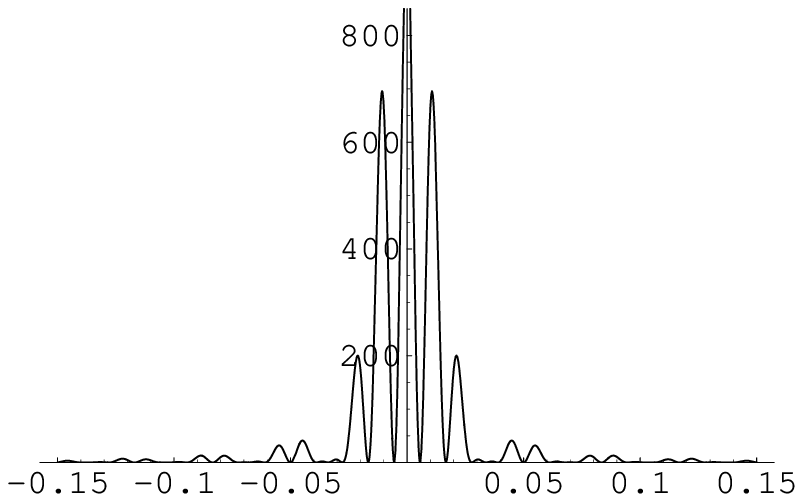}}}
 \put(10,20){\makebox(2,1)[l]{$I$}}
 \put(20,3.5){\makebox(2,1)[c]{$\beta(rad)$}}
\end{picture}
{\resizebox{5.5cm}{5.5cm}{\includegraphics{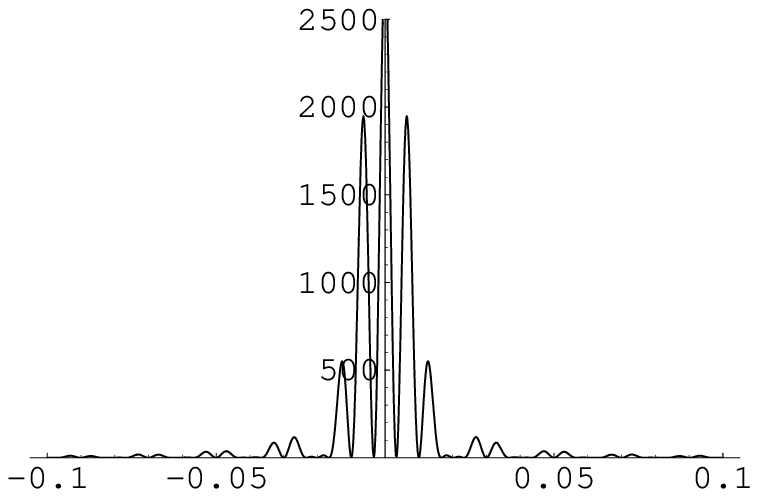}}} \vskip 5pt
{\hspace{0.1in}FIG. 9: The relation between $\beta$ and $I$
\hspace{0.35in} FIG.
10: The relation between $\beta$ and $I$ \\
 \hspace{0.3in} with $a=30\lambda$, $b=1000\lambda$, $c=\lambda$ and
$d=60\lambda$. \hspace{0.4in} with $a=50\lambda$, $b=1000\lambda$,
$c=\lambda$ and $d=100\lambda$.}
 \label{moment}
\end{figure}

\begin{figure}[tbp]
\begin{picture}(25,0)
 \put(6,20){\makebox(2,1)[l]{$I$}}
 \put(19,3.5){\makebox(2,1)[c]{$\beta(rad)$}}
{\resizebox{5.5cm}{5.5cm}{\includegraphics{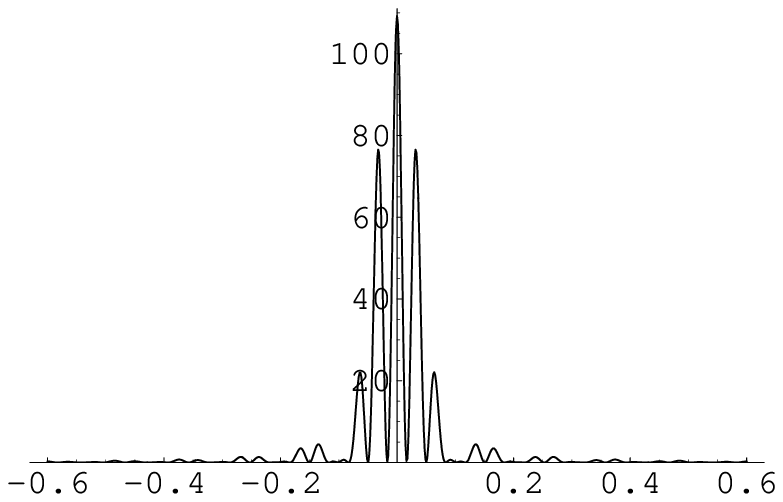}}}
 \put(10,20){\makebox(2,1)[l]{$I$}}
 \put(20,3.5){\makebox(2,1)[c]{$\beta(rad)$}}
\end{picture}
{\resizebox{5.5cm}{5.5cm}{\includegraphics{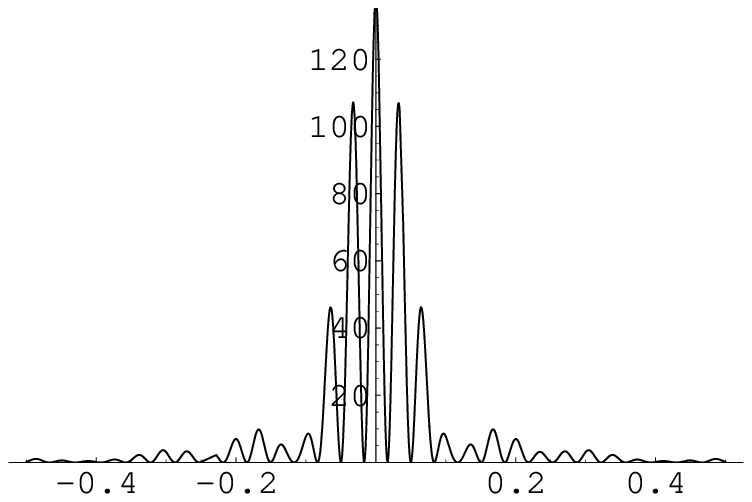}}} \vskip 5pt
{\hspace{0.1in}FIG. 11: The relation between $\beta$ and $I$
\hspace{0.35in} FIG.
12: The relation between $\beta$ and $I$ \\
 \hspace{0.3in} with $a=10\lambda$, $b=1000\lambda$, $c=0$ and $d=20\lambda$.
 \hspace{0.4in} with $a=10\lambda$, $b=1000\lambda$, $c=10\lambda$ and $d=20\lambda$.}
 \label{moment}
\end{figure}

\begin{figure}[tbp]
\begin{picture}(25,0)
 \put(6,20){\makebox(2,1)[l]{$I$}}
 \put(19,3.5){\makebox(2,1)[c]{$\beta(rad)$}}
{\resizebox{5.5cm}{5.5cm}{\includegraphics{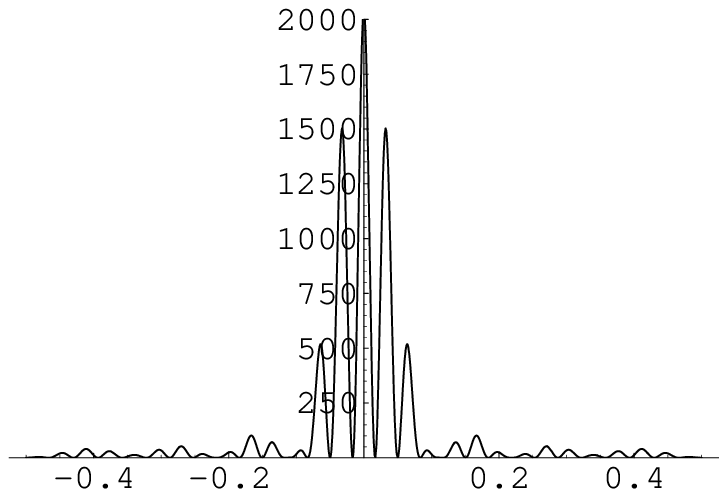}}}
 \put(10,20){\makebox(2,1)[l]{$I$}}
 \put(20,3.5){\makebox(2,1)[c]{$\beta(rad)$}}
\end{picture}
{\resizebox{5.5cm}{5.5cm}{\includegraphics{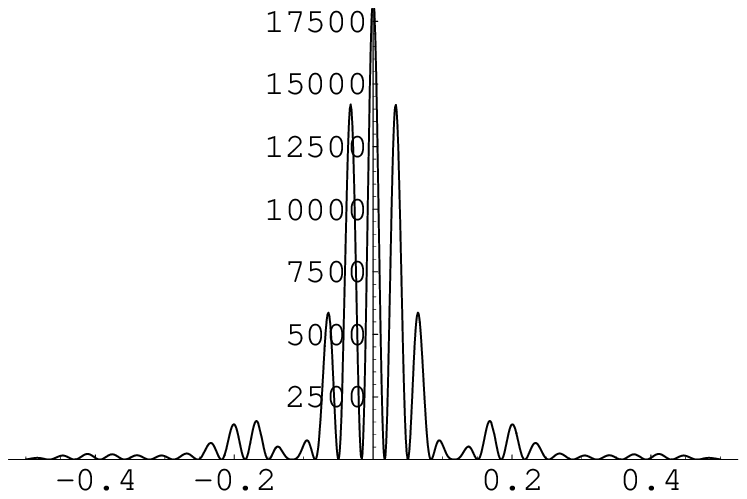}}} \vskip 5pt
{\hspace{0.1in}FIG. 13: The relation between $\beta$ and $I$
\hspace{0.35in} FIG.
14: The relation between $\beta$ and $I$ \\
 \hspace{0.3in} with $a=10\lambda$, $b=1000\lambda$, $c=100\lambda$ and $d=20\lambda$.
  \hspace{0.4in} with
$a=10\lambda$, $b=1000\lambda$, $c=1000\lambda$ and
$d=20\lambda$.}
 \label{moment}
\end{figure}


\begin{thebibliography}{10}

\bibitem{s1}
O. Carnal and J. Mlynek, Phys. Rev. Lett. {\bf66}, 2689 (1991).

\bibitem{s2}
W. Sch\"{o}llkopf, J. P. Toennies, Science {\bf266}, 1345 (1994).

\bibitem{s3}
M. Arudt, O. Nairz, J. Voss-Andreae, C. Kwller, G. Vander Zouw,
and A. Zeilinger, Nature {\bf401}, 680 (1999).

\bibitem{s4}
O. Nairz, M. Arudt and A. Zeilinger, J. Mod. Opt. {\bf47}, 2811
(2000).
\bibitem{s5}
S. Kunze, K. Dieckmann and G. Rempe, Phys. Rev. Lett. {\bf78},
2038 (1997).
\bibitem{s6}
B. Brezger, L. Hackermuller, S. Uttenthaler, J. Petschinka, M.
Arndt, A. Zeilinger, Phys. Rev. Lett. {\bf88}, 100404 (2002).
\bibitem{s7}
A.S. Sanz, F. Borondo and M.J. Bastiaans, Phys. Rev. A {\bf71},
042103 (2005).
\bibitem{s8}
M. Schwartz, Principles of Electrodynamics, Oxford University
Press, 1972.
\end{thebibliography}
\end{document}